\documentclass[aps,pra,groupedscriptaddress,showpacs]{revtex4}
\usepackage{setspace}
\usepackage[dvips]{graphicx}
\begin{document}
\begin{spacing}{2.0}
\title{Cooling of the rotation of a nanodiamond via the interaction with the electron spin of the contained NV-center}
\author{Li Ge}
\affiliation{School of Science, Hangzhou Dianzi University, Hangzhou, 310018, China}
\affiliation{Beijing Computational Science Research Center, Beijing, 100084, China}
\author{Nan Zhao}
\email{nzhao@csrc.ac.cn}
\affiliation{Beijing Computational Science Research Center, Beijing, 100084, China}

\begin{abstract}
   We propose a way to cool the rotation of a nanodiamond, which contains a NV-center and is levitated by an optical tweezer.
    Following the rotation of the particle, the NV-center electron spin experiences varying external fields and so leads
    to spin-rotation coupling. By optically pumping the electrons from a higher energy level to a lower level, the rotation
    energy is dissipated. We give the analytical result for the damping torque exerted on the nanodiamond, and
    evaluate the final cooling temperature by the fluctuation-dissipation theorem. It's shown that the quantum regime of
    the rotation can be reached with our scheme.

\end{abstract}

\pacs{37.10.vz, 37.30.+i, 42.50.Wk}

\maketitle

\section{Introduction}
Since the seminal experiments of Ashkin in 1970~\cite{ashkin1}, the techniques of optical trapping and manipulation have developed rapidly
over the past decades and stimulated remarkable advances in various fields of physics~\cite{ashkin2}. In atomic physics,
these techniques have greatly enhanced the ability to manipulate single atoms, leading to the experimental discovery of Bose-Einstein condensation~\cite{wieman, ketterle},
the implementation of atom interferometry~\cite{cronin} and quantum simulations of condensed-matter systems with cold atoms~\cite{bloch}.
More recently, optical manipulation has also been applied to larger objects such as micromirrors, cantilevers and dielectric nanoparticles to control the mechanical degrees of freedom~\cite{vahala,girvin1,favero,genes1,kiesel,schwab,roels,raizen}, with
the purposes of quantum information processing~\cite{tian,rips,barz}, ultrasensitive
  sensing~\cite{geraci1,geraci2,du,zhao} and studying quantum-classical boundaries~\cite{poot,chen}, etc.
  Theories regarding the cooling of center-of-mass (c.m.) motion were proposed~\cite{zwerger,zoller2,cirac1,cirac2}, and the
   quantum ground state cooling of a mechanical oscillator was realized experimentally~\cite{cleland,chan}. Besides,
   the interaction between the rotation of a nano-body and the light has also been investigated~\cite{cirac3,shi,hoang}.
   It is suggested that angular trapping and cooling of a dielectric can be achieved using multiple Laguerre-Gaussian
   cavity mode. The frequency of torsional vibration can be $1$ order of magnitude
   higher than the c.m. frequency~\cite{hoang}, which is promising for ground state cooling. Aside from being used
   for fundamental purposes, optically nanoparticle can also serve as ultrasensitive torque balance~\cite{kim,wu}.

   Recently, the coupling between the motion of a nanodiamond and the NV-center electron spin has attracted
   many research interests~\cite{xu,rabl,arcizet,lukin,yin}. The NV-center spin experiences varying external
   field following the motion of the nanodiamond, thus induces interaction between the spin and the mechanical
   motion, either translational or torsional.  In this paper we propose a cooling scheme based on the
   spin-rotation coupling, the mechanism of which is similar to that of atomic laser cooling~\cite{cohen1,cohen2}. A
   nanodiamond that contains a NV-center is levitated by an optical tweezer, hence its motion is confined.
    By applying external fields, the energy levels of NV-center are altered and left with
    an effective two-level system. In the course of rotation, the electrons in the higher level are optically pumped to the lower level,
    resulting in the dissipation of rotating energy of the nanodiamond, and thus achieves the effect of cooling.
    This paper is organized as folllowing: in Section I we outline the setup of the system and give a qualitative
    explanation of the cooling mechanism; Section II contains the calculation of the electronic state of the NV-center
    and the torque exerted on the nanodiamond; Based on these results, the cooling effect is analyzed in Section III;
    Finally we make conclusion in Section IV.


\subsection{Setup of the system}
We want to cool the rotation of a diamond nanoparticle that contains a NV-center.
Firstly, the center of mass motion and rotation of the particle should be confined, which can be achieved by an optical tweezer.
We consider a  spheroid shaped nanoparticle with semi-major axis $a$ and semi-minor axis $b$ placed in a linearly
polarized optical tweezer. If the size of the spheroid is much smaller than the
wavelength of the laser, the electrostatic approximation can be used to describe the light-matter interaction.
In a laser field with $Z$ polarization, the potential energy of the nanoparticle is:
\begin{equation}\label{trap}
  U=-\frac{1}{2}\sum_{i}\alpha_iE_i^2=-\frac{1}{2}\alpha_xE^2-\frac{1}{2}(\alpha_z-\alpha_x)E^2\cos^2\beta
\end{equation}
where $\alpha_x$ and $\alpha_z$ are the polarizibilities along the principal axes, $E$ is the electric field strength of the laser and $\beta$
is the nutation angle of the nanoparticle (Fig.1). The electric field at the laser waist is determined by:
 $\varepsilon_0E^2=\frac{2P}{\pi cw^2}$ , where $P$ is the laser power, $c$ is the light speed, and $w$ is the waist.
To give an estimation of the potential, we take $P=100mW$, $\pi w^2=2\mu m^2$, $a=40nm$, $b=20nm$, $\varepsilon_r=5.7$ (the dielectric constant of diamond), then 
the depth of the trap is
\begin{equation}
    U(\beta=\frac{\pi}{2})-U(\beta=0)\simeq 5.6\times10^{-21}J\equiv k_BT_0
\end{equation}
with $T_0\simeq 400K$.  At a temperature $T$, the nutation of the nanoparticle is confined in a range $[0, \beta_m]$, where
 $\beta_m$ is roughly determined by $k_BT_0(1-\cos^2\beta_m)=\frac{1}{2}k_BT$, e.g., for $T=5K$, $\beta_m\simeq 4.5^\circ$.

\begin{figure}
\begin{center}
\includegraphics[width=10cm]{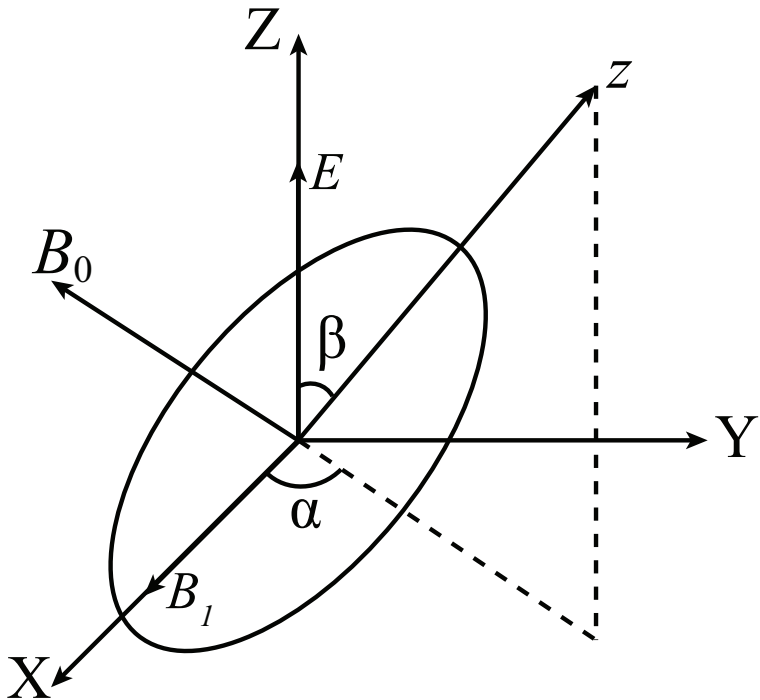} \label{Fig.1}
\caption{ \label{fig1} An ellipsoid rotates in external fields. $XYZ$ is the lab frame, $z$ is the major axis of the ellipsoid. $B_0$ is a static magnetic field that
lies in the $XZ$ plane and has an angle $45^\circ$  to the $X$ axis. There is also a laser field $E$ along the $Z$ axis and an oscillating magnetic field $B_1\cos\omega t$
along the $X$ axis. }
\end{center}
\end{figure}

\begin{figure}
\begin{center}
\includegraphics[width=10cm]{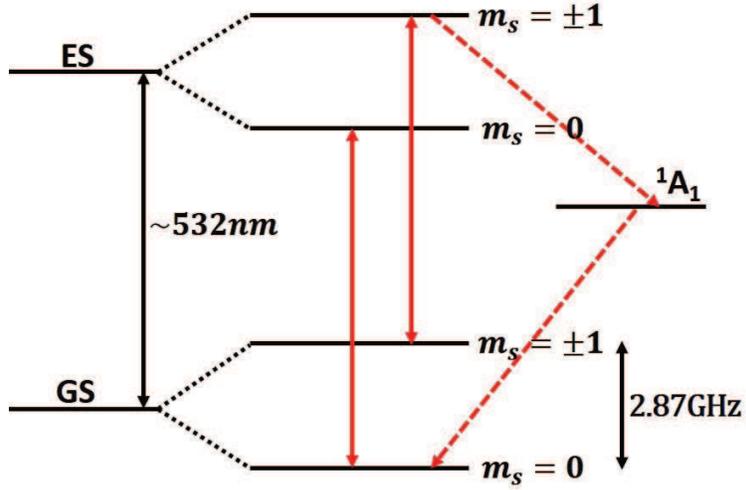} \label{Fig.2}
\caption{ \label{fig2} Electronic structure of the NV-center. GS is the ground state configuration and ES is the excited state configuration, each contains three
sublevels $m_s=0, \pm1$. A $532nm$ laser beam pumps the electrons in the $m_s=\pm1$ states in the GS to $m_s=0$ state. }
\end{center}
\end{figure}

Now that the trap is produced, the cooling shall be achieved through interactions between the NV center and external fields. The electronic states of  NV center
are illustrated in Fig.2, where $GS$ is the ground state configuration and $ES$ is the excited state configuration, each contains three spin sublevels with magnetic quantum number
$m_s=0, \pm1$. The sublevels in ground state have a zero-field splitting $D=2.87GHz$ between the states $|0\rangle$ and $|\pm1\rangle$.
Our cooling scheme borrows the idea from atomic cooling proposed by Cohen-Tannoudji \emph{et al} ~\cite{cohen1}. First, a static magnetic field $\mathbf{B}_0$ (Fig.1) is applied to modulate the energies
of states $|\pm1\rangle$
when the quantization axis of the NV center (which is the $z$ axis) rotates. This is analogous to the modulation of the light-shifted energies for a moving atom.
Second, another laser beam with $532nm$ wavelength is required to pump the electrons from $|\pm\rangle 1$ to $|0\rangle$ (Fig.2).
The pumping process includes spontaneous emissions to an intermediate state and then to the $|0\rangle$ state, which causes dissipation of energy. Finally we need a
microwave field $\mathbf{B}_1$ to induce transitions between $|0\rangle$ and $|\pm1\rangle$, since otherwise all the electrons shall be pumped
to $|0\rangle$ which is insensitive to external fields.

Now one may have a qualitative understanding of the cooling mechanism. If the frequency of the microwave field is $\omega$, the
effective energies of states $|\pm1\rangle$ are: $\delta_{\pm}=D-\omega\mp\gamma \mathbf{B}_0 \cdot \hat{z}$, where $\hat{z}$ is the unit vector along $z$ direction.
Since the rotation of the spheroid is confined by the optical trap, a suitable choice of $\omega$ and $\mathbf{B}_0$ makes that $\delta_{+}\ll\delta_{-}$
and the ground state reduces to a two level system with $m_s=0,1$. In the course of rotation, the electrons in level $|1\rangle$ are pumped to
$|0\rangle$, and since the state of this system doesn't follow the rotation adiabatically, as long as $\delta_{+}>0$ the energy is
continuously dissipated and a friction force is produced, otherwise the system absorbs more energy from the
$532nm$ laser than it loses in the spontaneous emissions. So to make sure the force is always frictional,
the rotation angle must be confined to a range in which $\delta_{+}>0$. In the following we give the details of the calculation,
which will confirm the above analysis.

\subsection{Calculation of the torque}
In the lab frame, the Hamiltonian describing the ground state configuration of the NV-center is:
\begin{equation}
H=D S_z^2-\gamma (\mathbf{B}_0+\mathbf{B}_1\cos \omega t)\cdot \mathbf{S}
\end{equation}
where $S_z$ is the $z$ component of the spin-1 operator. It relates to the spin operators in $XYZ$ frame as:
$S_z=\mathbf{S}\cdot\hat{z} =S_X\sin\beta\cos\alpha+ S_Y\sin\beta\sin\alpha+S_Z\cos\beta$.
 For a rotating nanoparticle, $\alpha, \beta$ and also $S_z$ are time dependent in the lab frame,
 making the calculations difficult. So it's better to move to the rotating frame, in which:
\begin{eqnarray}
H' &=& R^{\dagger}H R-iR^{\dagger}\frac{\partial R}{\partial t}=
D S_Z^2-\gamma (\mathbf{B}_0+\mathbf{B}_1\cos \omega t)\cdot R^{\dagger}\mathbf{S}R \nonumber \\
&=& D S_Z^2-\gamma (\mathbf{B}_0+\mathbf{B}_1\cos \omega t)\cdot R^{\dagger}\mathbf{S}R
-\dot{\alpha}(S_Z\cos\beta-S_X\sin\beta)-\dot{\beta}S_Y
\end{eqnarray}
where $R=e^{-i\alpha S_Z}e^{-i\beta S_Y}$ is the rotating operator. Physically, such transformation is analogous to the unitary transformation made in the rotating wave approximation.
In our setup, $\mathbf{B}_0$ is in the $XZ$ plane with an angle $45^\circ$ to the $X$ axis ($\mathbf{B}_0$ cannot be in the $Z$ direction.
The reason for this will be shown later), and $\mathbf{B}_1$ points to the $X$ direction.
This Hamiltonian is still time dependent due to the existence of microwave field $\mathbf{B_1}$, so we apply a
second unitary transformation: $U=e^{-i\omega t S_z^2}$, and make the rotating wave approximation,
then it becomes:
\begin{equation} \label{hamil}
H_e=(D-\omega) S_Z^2-\frac{1}{\sqrt{2}}\gamma B_0(\cos\alpha\sin\beta+\cos\beta)S_Z-\dot{\alpha}\cos\beta S_Z
-\frac{1}{2}\gamma B_1\cos\alpha \cos\beta S_X+\frac{1}{2}\gamma B_1\sin\alpha S_Y
\end{equation}
This Hamiltonian sets up the basis of our following calculations. The terms depending on $S_Z$ give the unperturbed energies of
$|0\rangle$ and $|\pm1\rangle$ , which are $E_0=0$ and $E_{\pm}=D-\omega\mp[\frac{1}{\sqrt{2}}\gamma B_0(\cos\alpha\sin\beta+\cos\beta)+\dot{\alpha}\cos\beta]$,
respectively. As discussed before, by suitably choosing $\omega$ and $B_0$ we can make $E_+\ll E_-$ so that this system becomes effectively two-level.
This Hamiltonian explicitly depends on the angles $\alpha$ and $\beta$, so the nanoparticle experiences a torque:
\begin{eqnarray} \label{torque}
\langle M_\alpha\rangle = \langle-\frac{\partial H_e}{\partial \alpha}\rangle=-\frac{1}{2}\gamma B_1\sin\alpha\cos\beta\langle S_x\rangle- \frac{1}{2}\gamma B_1\cos\alpha\langle S_y\rangle \nonumber \\
\langle M_\beta\rangle = \langle-\frac{\partial H_e}{\partial \beta}\rangle=\frac{1}{\sqrt{2}}\gamma B_0(\cos\alpha\cos\beta-\sin\beta)\langle S_z\rangle- \frac{1}{2}\gamma B_1\cos\alpha \sin\beta\langle S_x\rangle
\end{eqnarray}
where $\langle...\rangle$ means $Tr(\rho...)$ and $\rho$ is the density matrix describing the state of the NV-center.
In our setup, the microwave field $B_1$ is much smaller than the static field $B_0$, so in the following we focus on $M_\beta$.
The evolution of $\rho$  is governed by the master equation: $\dot{\rho}=-i[H_e, \rho]+\mathcal{D}(\rho)$, or, in the component form:
\begin{eqnarray} \label{master}
\dot{\rho}_{11} &=& -\frac{1}{\sqrt{2}} g_2(\rho_{01}+\rho_{10})+\frac{i}{\sqrt{2}}g_1(\rho_{01}-\rho_{10})-\Gamma\rho_{11} \label{up} \nonumber\\
\dot{\rho}_{01} &=& i(\delta_{+}-\dot{\alpha}\cos\beta){\rho}_{01}+\frac{i}{\sqrt{2}}g_1(\rho_{11}-\rho_{00})+\frac{1}{\sqrt{2}}g_2(\rho_{11}-\rho_{00})-\Gamma_1\rho_{01}
\end{eqnarray}
where we have introduced the notations $g_1=\frac{1}{2}\gamma B_1\cos\alpha \cos\beta$, $g_2=\frac{1}{2}\gamma B_1\sin\alpha$,
$\delta_{+}=D-\omega-\frac{1}{\sqrt{2}}\gamma B_0(\cos\alpha\sin\beta+\cos\beta)$.
$\Gamma$ and $\Gamma_1$ are decay rates for $\rho_{11}$ and $\rho_{01}$ respectively.

If the nanoparticle is at rest, i.e. $\dot{\alpha}=\dot{\beta}=0$, the steady state of the NV-center is obtained by putting
$\dot{\rho}_{11}=\dot{\rho}_{01}=0$. We get the steady values: $\rho^s_{11}=\frac{1}{2+f_0}$,
$\rho^s_{01}=\frac{(ig_1+g_2)f_0}{\sqrt{2}(2+f_0)(\Gamma_1-i\delta_{+})}$ with
 $f_0=\frac{\Gamma(\Gamma_1^2+\delta_{+}^2)}{\Gamma_1(g_1^2+g_2^2)}$.
For a rotating nanoparticle, the steady state satisfies $\dot{\rho}=\dot{\alpha}\frac{\partial\rho}{\partial\alpha}+\dot{\beta}\frac{\partial\rho}{\partial\beta}$,
and Eq. (\ref{master}) can be solved order by order. Note that $\Gamma_1\gg\Gamma$~\cite{liu}, so
 $\rho_{01}$ follows almost adiabatically with the evolution of $\rho_{11}$~\cite{cohen2}:
$\rho_{01}=\frac{(ig_1+g_2)(\rho_{11}-\rho_{00})}{\sqrt{2}(\Gamma_1-i\delta_{+})}$. Inserting this formula to the first equation of (\ref{master}) we get:
\begin{equation}
\dot{\alpha}\frac{\partial\rho_{11}}{\partial\alpha}+\dot{\beta}\frac{\partial\rho_{11}}{\partial\beta}=-\frac{(2+f)\Gamma}{f}(\rho_{11}-p_s)
\end{equation}
where $f=\frac{\Gamma[\Gamma_1^2+(\delta_{+}-\dot{\alpha}\cos\beta)^2]}{\Gamma_1(g_1^2+g_2^2)}$, $p_s=\frac{1}{2+f_0}$.
Expanding $\rho_{11}=p_s+p^{(1)}+...$ and $\frac{(2+f)\Gamma}{f}=\frac{(2+f_0)\Gamma}{f_0}+\dot{\alpha}A_1+...$,
then to the first order of $\dot{\alpha},\dot{\beta}$:
\begin{equation}
\dot{\alpha}(\frac{\partial p_s}{\partial\alpha}-A_1p_s)+\dot{\beta}\frac{\partial p_s}{\partial\beta}=-\frac{(2+f_0)\Gamma}{f_0}p^{(1)}
\end{equation}
Hence
 \begin{equation}  \label{linear}
   p^{(1)}=-\frac{f_0}{\Gamma(2+f_0)}\big(\frac{\partial p_s}{\partial\alpha}\dot{\alpha}-A_1p_s\dot{\alpha}
    +\frac{\partial p_s}{\partial\beta}\dot{\beta}\big)
 \end{equation}

And from Eq. (\ref{torque}):
\begin{equation}
\langle M_{\beta}\rangle\approx\frac{1}{\sqrt{2}}\gamma B_0(\cos\alpha\cos\beta-\sin\beta)\big(p_s-\frac{f_0}{\Gamma(2+f_0)}\frac{\partial p_s}{\partial\alpha}\dot{\alpha}
-\frac{f_0}{\Gamma(2+f_0)}\frac{\partial p_s}{\partial\beta}\dot{\beta}\big)\equiv M_{\beta}^0-\kappa_{\alpha\beta}\dot{\alpha}-\kappa_{\beta}\dot{\beta}
\end{equation}
where $ M_{\beta}^0$ is the conservative force and:
\begin{eqnarray} \label{coeff}
\kappa_{\alpha\beta}&=&\frac{1}{\sqrt{2}}\gamma B_0(\cos\alpha\cos\beta-\sin\beta)\frac{f_0}{\Gamma(2+f_0)}\frac{\partial p_s}{\partial\alpha}
\simeq-\frac{(\gamma B_0)^2\delta_{+}f_0}{\Gamma_1(g_1^2+g_2^2)(2+f_0)^3}\sin\alpha\sin\beta(\cos\alpha\cos\beta-\sin\beta) \nonumber  \\
\kappa_{\beta}&=&\frac{1}{\sqrt{2}}\gamma B_0(\cos\alpha\cos\beta-\sin\beta)\frac{f_0}{\Gamma(2+f_0)}\frac{\partial p_s}{\partial\beta}
\simeq[\gamma B_0(\cos\alpha\cos\beta-\sin\beta)]^2\frac{\delta_{+}f_0}{\Gamma_1(g_1^2+g_2^2)(2+f_0)^3}
\end{eqnarray}

\begin{figure}
\begin{center}
\includegraphics[width=10cm]{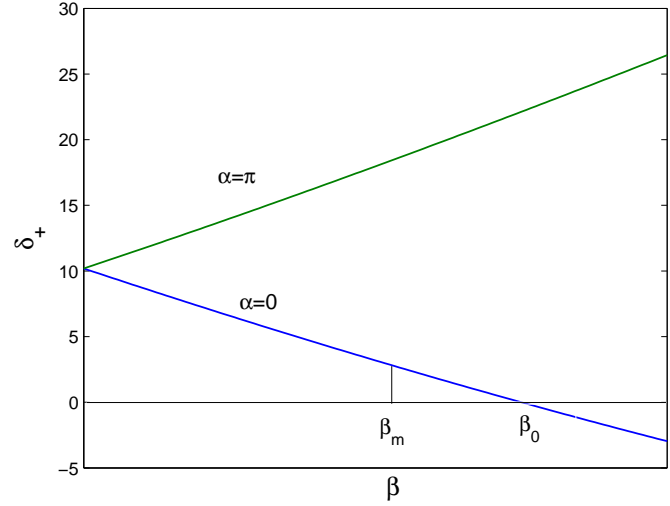} \label{Fig.3}
\caption{\label{fig3} A sketch of $\delta_{+}$ as a function of $\beta$. The upper curve is for $\alpha=\pi$ and lower curve for $\alpha=0$,
the value of $\delta_{+}$ for arbitrary $\alpha$ is between these two curves. At $\beta=\beta_0$ the lower curve crosses zero, so as long as
$\beta\leq\beta_m<\beta_0$, $\delta_+$ is always positive. }
\end{center}
\end{figure}
The effect of $\kappa_{\alpha\beta}\dot{\alpha}$ will be discussed later and we focus on $\kappa_{\beta}\dot{\beta}$ now, which is a friction force
as long as $\kappa_{\beta}>0$. From the above equation it's seen that the sign of $\kappa_{\beta}$ is determined by $\delta_{+}$, which confirms the qualitative analysis in the previous
section. As we mentioned before $\beta$ is confined by the optical trap in a range $[0, \beta_m]$, then by suitably choosing the frequency $\omega$ such that at a certain angle
$\beta_0>\beta_m$ (but smaller than $45^\circ$), $D-\omega-\gamma B_0\cos(45^\circ-\beta_0)=0$, one can make sure $\delta_{+}$ is always positive
in the course of rotation (Fig.3).

The reason for that $\mathbf{B}_0$ can't lie in the $Z$ axis is now clear. It's seen $\kappa_\beta$ contains a factor $[B_0(\cos\alpha\cos\beta-\sin\beta)]^2$
which is proportional to $[\frac{d}{d\beta}(\mathbf{B}_0\cdot \hat{z})]^2$, then if $\mathbf{B}_0$ is in the $Z$ axis this will be $(B_0\sin\beta)^2$, making
the friction force negligible for small $\beta$.

\subsection{Analysis of cooling effect}

We are now going to estimate the time scale over which the rotation is damped and the final cooling temperature.
For small $\beta$, the Hamiltonian for the nanoparticle, which consists of the rotating energy and the potential of optical trap is:
\begin{equation}
  H_p=\frac{I_1}{2}(\dot{\beta}^2+\dot{\alpha}^2\sin^2\beta)+\frac{I_3}{2}(\dot{\alpha}\cos\beta+\dot{\gamma})^2+U
  \simeq\frac{I_1}{2}\dot{\beta}^2+\frac{I_3}{2}(\dot{\alpha}+\dot{\gamma})^2+\frac{1}{2}(\alpha_z-\alpha_x)E^2\beta^2
\end{equation}
where a constant energy is omitted, and $I_1$, $I_3$ are moments of inertial about the $x$ and $z$ axes respectively. The torque exerted on the
particle is produced by the optical trap as well as $M_\beta$, so the motion equation of $\beta$ is:
\begin{equation}\label{beta}
  I_1\ddot{\beta}=-\frac{\partial U}{\partial \beta}+M_{\beta}=-\frac{\partial U}{\partial \beta}+M_{\beta}^0-\kappa_{\alpha\beta}\dot{\alpha}-\kappa_{\beta}\dot{\beta}+M'_{\beta}
\end{equation}
 where $M'_{\beta}=M_\beta-\langle M_\beta\rangle$ is the fluctuation of the torque.
 The time scale for damping is $t=\frac{I_1}{\kappa_{\beta}}$. In our case, $I_1=\frac{4\pi}{15}\rho ab^2(a^2+b^2)$, where $\rho=3.5g/cm^3$ is the density of diamond,
 the other parameters are : $\Gamma=0.4\mu s^{-1}/2\pi$, $\Gamma_1=5\mu s^{-1}/2\pi$ and $\beta_0=9^\circ$.
 The dependence of $\kappa_\beta$ on $\beta$ is shown in Fig.4, where we take $\alpha=0$ and $\pi$ for examples.
To give a rough estimation, taking $\kappa_\beta=500h$ for average, then $t\simeq 2.8\times10^{-4}s$.

\begin{figure}
\begin{center}
\includegraphics[width=10cm]{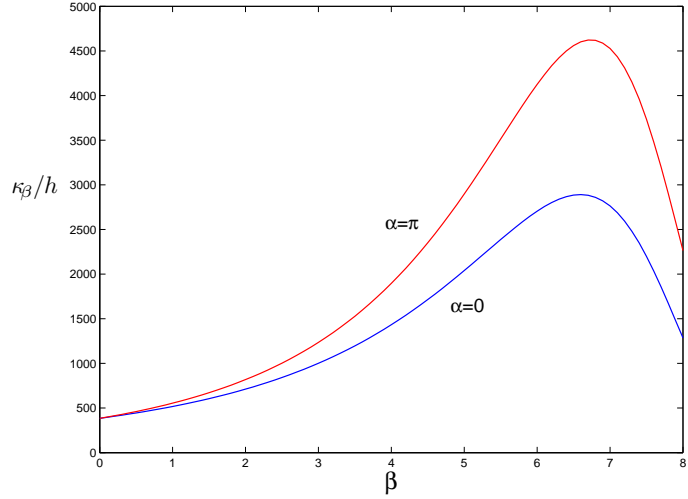} \label{Fig.4}
\caption{\label{fig4} The $\beta$ dependence of $\kappa_{\beta}$ (in unit of the Planck constant $h$) with the following parameters: $\gamma B_0=2\pi\times100MHz$, $\gamma B_1=2\pi MHz$,
$\Gamma=0.4 MHz$, $\Gamma_1=5MHz$ and $\beta_0=9^\circ$. The upper curve is for $\alpha=\pi$ and lower curve for $\alpha=0$.}
\end{center}
\end{figure}


The damping of $\dot{\alpha}$  is much slower than that of $\dot{\beta}$ since from Eq.(\ref{torque}) and (\ref{linear}),  $M_{\alpha}\propto \gamma B_1$ ,
and is negligible compared to $M_\beta$ for $B_1\ll B_0$. So after some time of cooling, $\dot{\alpha}\gg\dot{\beta}$ and in a period $\tau$ that $\alpha$
changes from $0$ to $2\pi$, $\beta$ can be taken as a constant. Then the impulse of $\kappa_{\alpha\beta}\dot{\alpha}$ in such a period is:
\begin{equation}
    \int_0^{\tau}\kappa_{\alpha\beta}\dot{\alpha}dt=\int_0^{2\pi}\kappa_{\alpha\beta}d\alpha=0
\end{equation}
where the last equality results from $\kappa_{\alpha\beta}(\alpha)=-\kappa_{\alpha\beta}(2\pi-\alpha)$.


To estimate the final cooling temperature, we need to calculate the correlation function of the friction force. Let $G(t)=\langle S_z(t)S_z(0)\rangle-\langle S_z\rangle^2$,
since $\langle S_z\rangle= p_s$, we have from the quantum regression theorem~\cite{breuer} that $\frac{d G}{dt}=-\frac{(2+f_0)\Gamma}{f_0}G$ with initial condition
$G(0)=p_s-p_s^2$. Then $G=G(0)\exp(-\frac{(2+f_0)\Gamma}{f_0}t)$ and, from (\ref{torque}):
\begin{equation}
\langle M'_\beta(t)M'_\beta(0)\rangle = \frac{1}{2}[\gamma B_0(\cos\alpha\cos\beta-\sin\beta)]^2(p_s-p_s^2) \exp(-\frac{(2+f_0)\Gamma}{f}t)
\end{equation}
Hence the momentum diffusion coefficient is:
\begin{equation}
D_p=\frac{1}{2}\int_{-\infty}^{\infty}\langle M'_\beta(t)M'_\beta(0)\rangle dt= \frac{1}{2}[\gamma B_0(\cos\alpha\cos\beta-\sin\beta)]^2(p_s-p_s^2) \frac{f_0}{(2+f_0)\Gamma}
\end{equation}
where we have used the fact that for $t<0$, $\langle S_z(t)S_z(0)\rangle=\langle S_z(-t)S_z(0)\rangle$.
So according to the fluctuation-dissipation theorem~\cite{breuer}, the temperature is:
 \begin{equation}  \label{tem1}
 k_BT_f=\frac{D_p}{\kappa_\beta}=\frac{(2+f_0)^2 (p_s-p_s^2)(g_1^2+g_2^2)\Gamma_1}{\delta_+\Gamma}=\frac{(1+f_0)(g_1^2+g_2^2)\Gamma_1}{\delta_+\Gamma}.
 \end{equation}
The angle $\beta$ will be about $0$ after a period of cooling, so we take $\beta=0$ in the above equation.
Then $\delta_+=\gamma B_0[\cos(45^\circ-\beta_0)-\cos45^\circ]$ and $g_1^2+g_2^2=(\frac{1}{2} \gamma B_1)^2$. Let's
consider the case $f_0=\frac{\Gamma(\Gamma_1^2+\delta_{+}^2)}{\Gamma_1(g_1^2+g_2^2)}\gg1$, which can be fulfilled
for $\gamma B_0\gg\gamma B_1$, then Eq. (\ref{tem2}) is simplified as:
 \begin{equation}  \label{tem2}
 k_BT_f\simeq\frac{f_0(g_1^2+g_2^2)\Gamma_1}{\delta_+\Gamma}=\frac{\Gamma_1^2+\delta_+^2}{\delta_+}
 \end{equation}
So the temperature reaches its minimum $2\Gamma_1/k_B$ at $\delta_+=\Gamma_1$.  For $\Gamma_1=5MHz $, the lowest
temperature is $T_f\simeq 0.6\times10^{-4}K$.  We want to see whether this temperature is low enough to reach the quantum regime.
The Hamiltonian $H_p$ describes a harmonic oscillator with frequency $\omega_0=\sqrt{(\alpha_z-\alpha_x)E^2/I_1}$, so the
 temperature for quantum regime is $\hbar\omega_0/k_B\sim 10^{-4}K$, which is of the same order of $T_f$.

\section{Conclusion}
To conclude, we study the rotation cooling of a nanodiamond which contains a NV-center.
 Through the coupling between its rotation and the NV-center electron spin,
the rotation energy is dissipated, which is similar to the atomic laser cooling. By suitably choosing the parameters of
the system setup, the quantum regime can be reached. In our theory, the motion of nanoparticle is treated classically.
However when the temperature is low enough, quantum effect must be taken into account.
A full quantum mechanical approach is of our future interest.

Above we have assumed that there is only one NV-center
in the nanodiamond. If the number of the NV-center is $n$, and if these NV-centers are uncorrelated,
the damping coefficient $\kappa_\beta$ should be $n$ times the present one. However the final cooling temperature
is unchanged since $D_p$ is also multiplied by a factor $n$.

 \section{acknowledgements}
  This work is supported by Natural Science Foundation of Zhejiang Province LQ18A040003, NKBRP (973 Program) 2014CB848700 and No. 2016YFA0301201, NSFC No. 11534002 and NSAF U1530401, Science Challenge Project No.TZ2018003.

\end{spacing}
\end{document}